# A Heuristic EDF Uplink Scheduler for Real Time Application in WiMAX Communication

**Nidhi Lal, Anurag Prakash Singh, Shishupal Kumar, Shikha Mittal, Meenakshi Singh**
M.Tech in Wireless Communication and Computing,
IIIT-Allahabad, India

***Abstract*— WiMAX, Worldwide Interoperability for Microwave Access, is a developing wireless communication scheme that can provide broadband access to large-scale coverage. WiMAX belongs to the family of standards of IEEE-802.16. To satisfy user demands and support a new set of real time services and applications, a realistic and dynamic resource allocation algorithm is mandatory. One of the most efficient algorithm is EDF (earliest deadline first). But the problem is that when the difference between deadlines is large enough, then lower priority queues have to starve. So in this paper, we present a heuristic earliest deadline first (H-EDF) approach of the uplink scheduler of the WiMAX real time system. This H-EDF presents a way for efficient allocation of the bandwidth for uplink, so that bandwidth utilization is proper and appropriate fairness is provided to the system. We use Opnet simulator for implementing the WiMAX network, which uses this H-EDF scheduling algorithm. We will analysis the performance of the H-EDF algorithm in consideration with throughput as well as involvement of delay.***

## I. Introduction

WiMAX inherits IEEE 802.16 features which are operating in the 10 to 66 GHz range. [1] WiMAX has a property of interoperability which means that it supports various types of wireless device. WiMAX supports a very wide range of mobility so that nodes of WiMAX system can freely move to one place to another place. It also supports handoff (soft handover and hard handover). It uses scalable orthogonal frequency-division multiple access (SOFDMA) with 256 sub-carriers in place of the fixed orthogonal frequency-division multiple access. [2, 16] With a rapid growth of Broadband Wireless Access, WiMAX became very interesting and famous topic in the research area as well as in the field of industry during the past few years. To sustenance the QOS (Quality of Service) prerequisite, the IEEE 802.16 standard postulates five scheduling classes of services: Unsolicited Grant Service (UGS), extended real-time Polling Service (ertPS), real-time Polling Service (rtPS), non-real time Polling Service (nrtPS), and Best Effort (BE). [3] To achieve the efficient and reliable allocation of in WiMAX networks, we need a scheduling algorithm. The scheduling algorithm is a substantial portion of QOS architecture. There are two types of algorithms: downlink algorithm (from BS to SSs), and uplink algorithm (from SSs to BS). The MAC layer of the WiMAX structure utilizes a scheduling algorithm because the subscriber station needs to compete for initial entry into the network. After making an entry in the network, an access slot allocated by the base station to the corresponding subscriber station. The assigned time slot to the subscriber station is unique i.e. it cannot be used by another subscriber station in the WiMAX network. The scheduling algorithm does the efficient management of assignment of time slots to subscriber station and by this way it can properly control and balance the parameters Quality of Service (QoS). Figure 1 shows a simple architecture of WiMAX communication system.

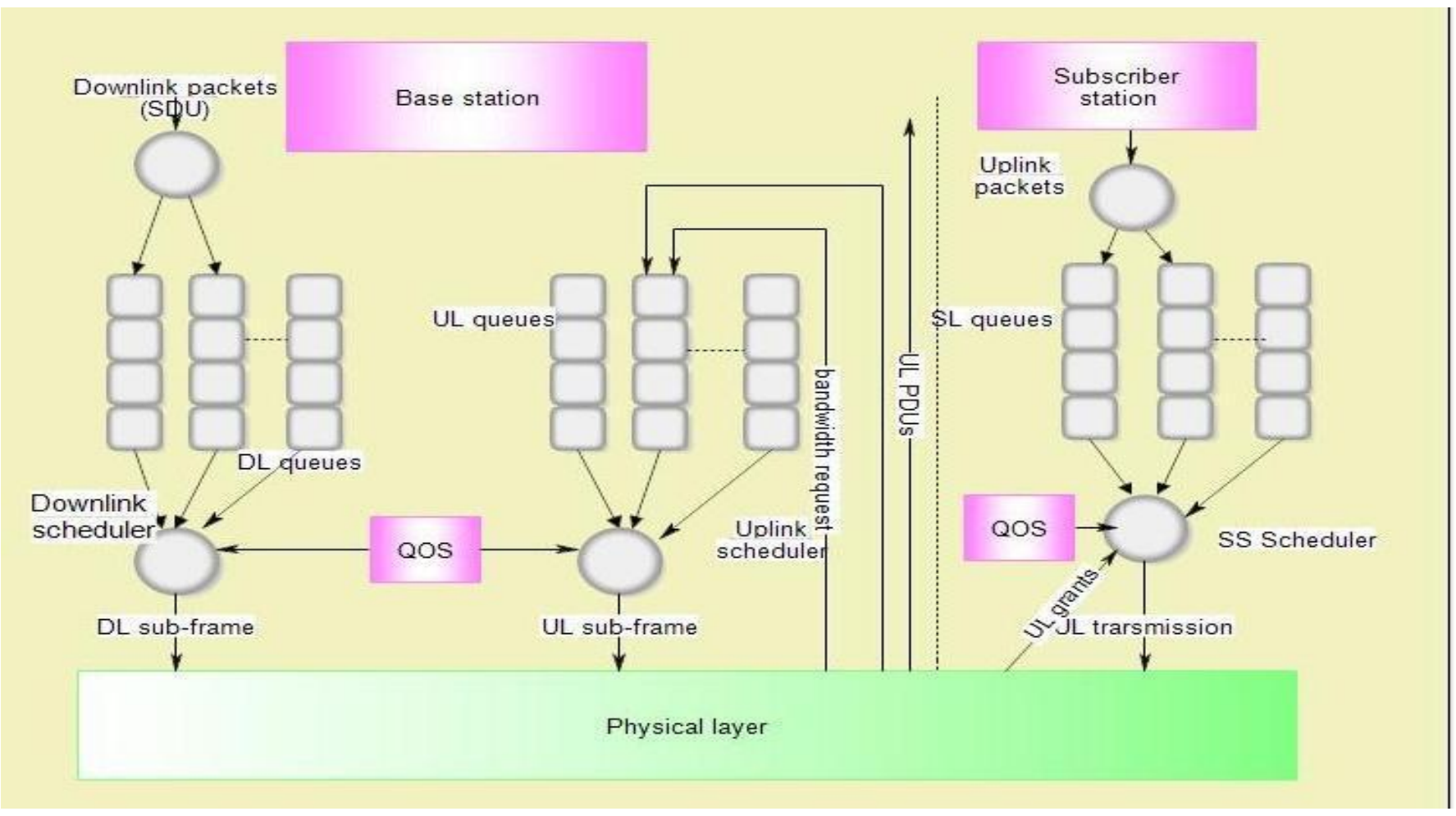

Fig 1: Simple layout of WiMAX communication system

Over the past few years, there has been a speedy development and progress of the new application area and services such as online video games, sports, video conferencing and so many multimedia services which belongs to real time application. To manage and control of QoS requirement of these real time applications and services has now become a major concern because it does great impact on the performance of WiMAX network and these real time applications need high data speed. To balance the QoS requirements of these real time multimedia applications, a strong scheduling algorithm is needed which efficiently and reliably assign the bandwidth to users such that maximum throughput achieves with low end to end delay. The main concern of the scheduling algorithm is done by dispensing the bandwidth and resources amid subscriber stations (SSs). Some of the works had been done on the topic of scheduling algorithm in WiMAX. Those algorithms are well known by the name of RR (Round Robin) and WRR (Weighted Round Robin) scheduling algorithms. The first one, the Round Robin algorithm can be painstaking the very first meek algorithm which basically assigns the bandwidth by selecting one connection in all connections. Moreover, RR contains some disadvantages like it allocates bandwidth to the subscriber station that may have nothing to transmit and this phenomena leads to wastages of bandwidth resources. Meanwhile RR cannot give assurance to provide a good QoS for dissimilar service classes in WiMAX communication scheme. The second one Weighted Round Robin (WRR) has been granted for WiMAX scheduling. In this scheduling algorithm, the weights can be used to provide fine-tuning between throughput and delay necessities. WRR is an extension lead of the Round Robin algorithm. To overcome the limitations of the above algorithm, Earliest Deadline First (EDF) scheduling algorithm is proposed which works for uplink real time polling services (rtPS) packets. EDF leads to starvation problem and also not support high mobility. So one another algorithm is introduced in WiMAX communication network that is called proportional fairness (PF). PF works on by assigning priority to the tasks but it does not provide efficient quality of service. [4]So an improved version of EDF is implemented which is based on the hybrid model and uses historical throughput to calculate the priority of the nodes included in the bandwidth request which is called Subscriber Stations Based Proportional Fairness(SSBPF) scheduler which is based on the Grant Per Type-of-Service (GPTS) principle and provide QoS to real time applications and services.

## II. RELATED WORK

Scheduling for real time application and multimedia application comes under real time polling services (rtPS) which is not defined in IEEE 802.16 standard. [4] IEEE 802.16 has comprehensive papers about the QoS parameter. However, they do not declare the scheduling algorithm to be used. To perfectly balance the QoS parameter and provide a fair allocation of bandwidth among the subscriber station by the base station is a major concern of the scheduling algorithm used in WiMAX communication system. Certain algorithms for WiMAX are [5] Round Robin (RR) is a commonly used scheduling algorithm in WiMAX with QoS control, where users are scheduled at regular intervals with assignment of equal service chance. [6] The other one is weighted fair queue (WFQ), which is an estimation of General Processor Sharing (GPS), in which each connection is assigned to its individual FIFO queue and the weight can be vigorously consigned for each queue. The sharing of the available resources is according to the proportion of the assigned weight. [7] Weighted round-robin (WRR) is used as an approximation to WFQ having aim to provide some gradation of separation between queues. [8] Largest weighted delay first scheduling aims avoidance of missing a deadline so that it chooses the packet which having largest delay. [9] Delay Threshold Priority Queuing (DTPQ) was proposed for usage when in cooperation real-time and non-real-time traffic in chorus are contemporary. In this scheduling algorithm, a higher priority is given to real-time traffic but that event could harm the non-real time traffic. To overcome this issue a condition is defined. According to this condition, the real-time traffic is assigned to bandwidth only when the head-of-line (HOL) packet delay surpasses a specified delay threshold. [10]Deficit Fair Priority Queuing (DFPQ) by way of using a counter was introduced to balance and control of allocation maximum allowable bandwidth for each service class. The counter diminutions rendering to the size of the packets. The scheduler vagaries to alternative class when the counter value condensed to zero. [11] Modified Largest Weighted Delay First (M-LWDF) can offer better QoS pledges but it gives a reduced amount of throughput and also it provides guarantee to keep static the value of delay smaller than a predefined threshold. [12] To maintain the delay bound i.e. it always less than the predefined threshold value and to provide some fairness in the system in harsh conditions, a minor alteration of the WFQ is proposed which is called Worst-case fair Weighted Fair Queuing (WF2Q). This scheduling algorithm is Similar to WFQ, but WF2Q uses a virtual time perception. The virtual finish time is the time that a GPS would have complete sending the packet. WF2Q, selects a packet which having smallest virtual finishing time and whose virtual start time has already occurred.

## III. EDF ALGORITHM FOR WIMAX COMMUNICATION SYSTEM

There are many scheduling algorithms are proposed for uplink layer of the WiMAX. One of the most efficient algorithm is EDF (earliest deadline first). This algorithm gives very high performance and provides fairness to the system. EDF is basically used for real time system packets (rtPS) in WiMAX. We first have to implement a basic EDF algorithm, Earliest deadline first (EDF) or less time to go is a dynamic scheduling algorithm used in real-time operating systems to place a bandwidth request of subscriber station in a priority queue. Whenever a new bandwidth request by a subscriber station occurs then the queue will be rifled for the process which is contiguous to its deadline. This selected process is one who has to be scheduled succeeding for execution. In EDF priority can be provided obliquely by assigning dissimilar static deadlines for different traffic classes, the corresponding requests have smaller deadline will be scheduled earlier than requests having higher deadline. But the problem arises in that case when the difference among those requests deadlines is large enough. In such a situation, lower priority request have to wait for very long time and this phenomena leads to starvation of that subscriber station. So, EDF leads to starvation problem and also not support high

mobility.so one another algorithm is introduced in WiMAX communication network that is called proportional fairness (PF). [14-15]PF mechanism on by allocating priority to the tasks but it was not able to provide a balanced and efficient quality of service. So to overcome these problems an algorithm of EDF scheduling is modified in a certain way, which is based on the hybrid model. [4]In this model, it is observed for the historical throughput of each requesting subscriber station and after that it calculates priority of the nodes. These overall procedure is known as Subscriber Stations Based Proportional Fairness (SSBPF) scheduler.

## IV. PROPOSED WORK

In this paper, our goal is to implement a uplink scheduler for WiMAX communication which schedules the bandwidth request of subscriber station(SS) in an efficient way by using earliest dead line first (EDF) real time application algorithm with some modifications so that it always allocates bandwidth to the subscriber station in efficient way and provide fairness to the system. The work belongs to implementing an uplink scheduler which we call heuristic earliest deadline first (H_EDF) uplink scheduler for real time application in WiMAX communication uses a heuristic approach which will overcome the limitations of EDF algorithm and much faster than EDF. This scheduling algorithm provides a better utilization of the bandwidth and provide a fair allocation of bandwidth to the subscriber station. As well as, it protects the uplink scheduler from the waste of bandwidth and by this way it decreases the overall communication delay. It provides a secure communication network between the sender node to the receiver node. It minimizes the packet loss rate. In this paper, we first implement an improved EDF scheduling algorithm using [4]SSBPF scheduling approach and will measure its performance in terms of packet loss, end to end delay and throughput. Further, we Implement H-EDF scheduling algorithm which is based on [13] heuristic function in the WiMAX environment. To avoid Starvation, the priority of the mobile node subscriber station (SS) is calculated with the help of historical throughput. The subscriber station, which has more historical throughput have higher the priority and according to this bandwidth is allocated by the base station to the corresponding subscriber station. And then we will compare and analysis the performance of the SSBPF scheduling algorithm with implementing the H-EDF algorithm. Figure 2 shows the framework of our proposed H-EDF scheduling algorithm in WiMAX communication scheme.

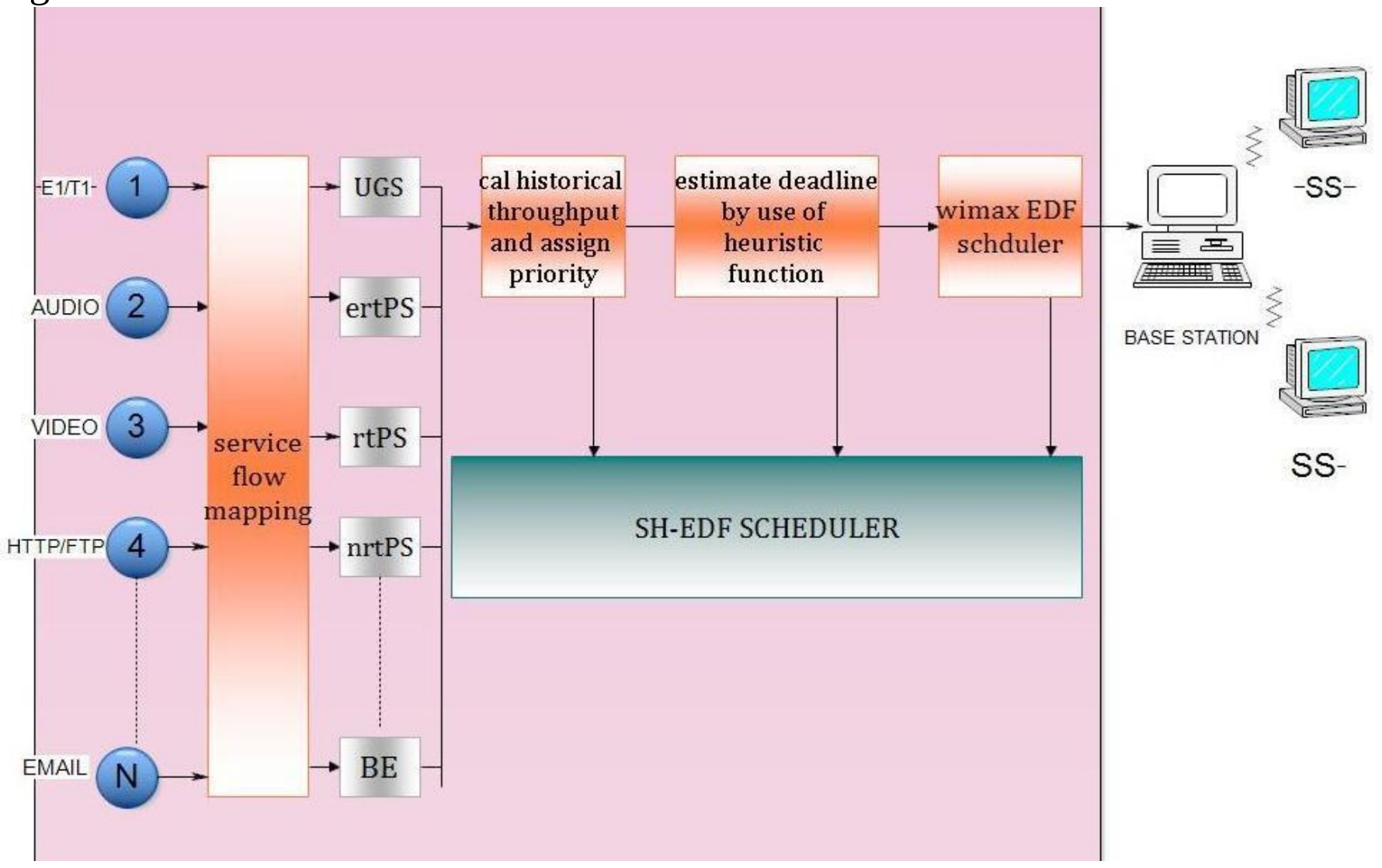

Fig 2: Framework of proposed H-EDF uplink scheduler

To avoid the limitations of EDF algorithm, we use a heuristic mathematical approach, [13] which uses heuristic function to calculate the deadline and with the help of this it determines which process should be executed next. In this way, the allocation of bandwidth will be in a fair and good manner and it leads to high performance of the system of WiMAX communication. The following steps are involved while determining the process which will schedule next by using heuristic mathematical model.

- To avoid the starvation problem in the EDF scheduling algorithm, [4] an approach of the SSBPF scheduling algorithm is used here. In this SSBPF algorithm, each task is assigned a priority. This priority is calculated based on the previous throughput value by using the following formula:

  Priority (i) = c(i)/(1+th(i))

Where th(i) denotes historical throughput and c(i) denotes transmission capacity. It is shown that when a subscriber station node have more historical throughput then higher priority is given to that node. This method leads to utilization of bandwidth in more enhanced and fair way and leads to high performance of WiMAX communication system. For further enhancing the performance of WiMAX, a heuristic function is used for determining the node which has to execute next. [13]The approach is described in the following steps:

- If new task enter into queue, comparisons starts by calculating 'Claim' value μ(t) by using following formula: μ(t)=total execution time of next task(burst time) + remaining execution time of current task + current time
  Remaining execution time=total execution time – current time

- The decision of Next executed Task N(T) will be taken by comparing μ (t) value of current process with D(j) which is the deadline of next process.
- [N(t)]= { current task( μ (t)) <=D (j) , next executed task( μ (t))> D (j)}

H-EDF will provide high throughput and hence it will increase the performance of WiMAX communication system with less wastage of bandwidth. H-EDF uplink scheduling algorithm have very less context switches and it is very useful for real time application in WiMAX communication system. Real time applications are coming under the Silver zone of WiMAX [18] supported services like video conferencing and it requires more bandwidth [19] to transfer the data. Our proposed H-EDF uplink scheduler will give better results and very useful for real time communication system.

## V. SIMULATION PARAMETERS

In this paper, the simulation of proposed H-EDF uplink scheduler will be on OPNET Modeler 14.5 simulator. This simulator requires knowledge of C and C++ languages. This simulator runs on Windows 7 ultimate 32 bit operating system, Intel core. It requires installation of visual studio for running the simulation works of OPNET simulator. A simple architecture of WiMAX communication system using OPNET simulator is shown in the figure 3:

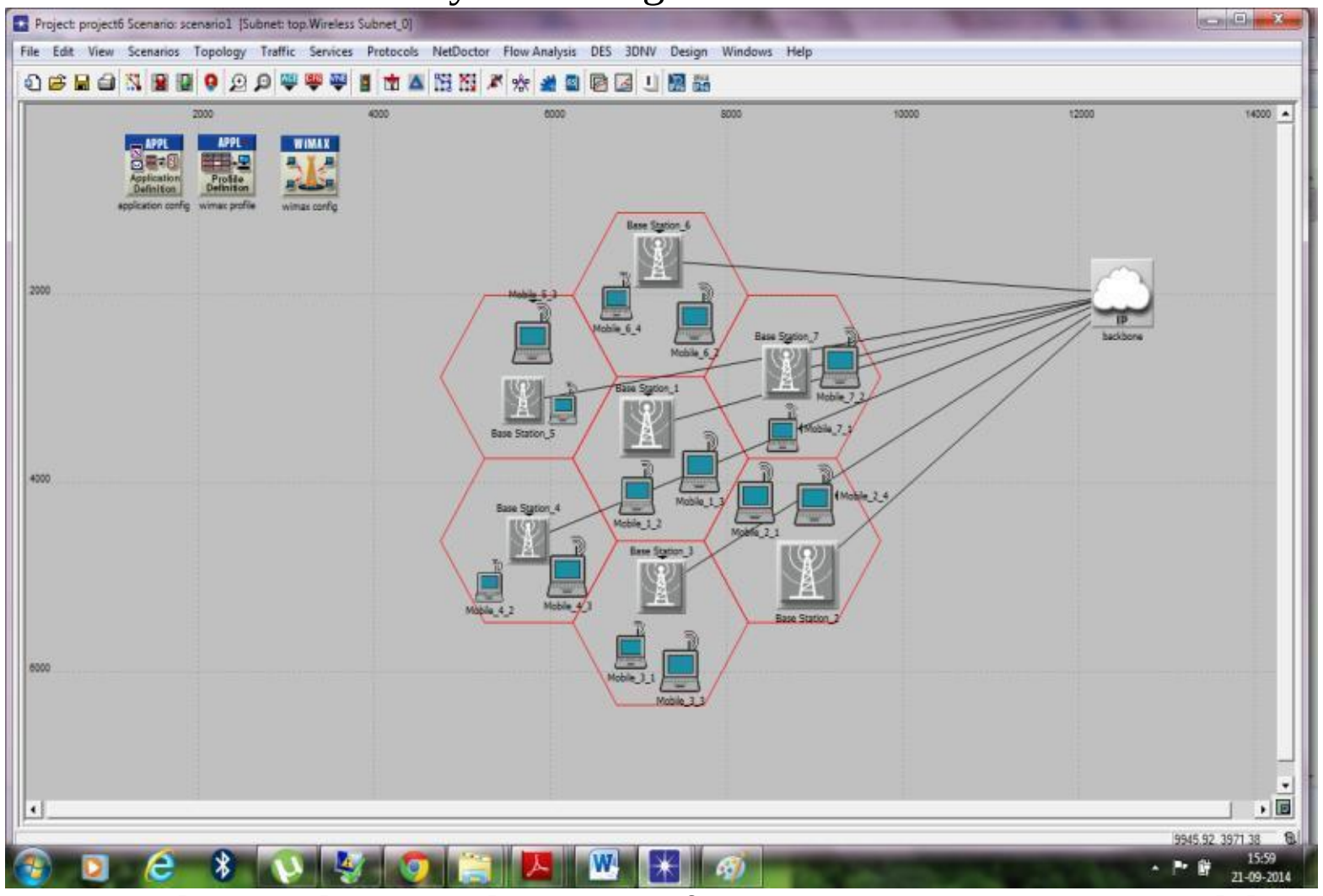

Fig 3: The simulation architecture of WiMAX using OPNET Modeler

In above figure 3 shows a simulation of WiMAX communication system. In our simulated scenario 7 hexagonal cells are taken. In every single cell, each contain 2 subscriber stations and have only 1 base station. The simulated scenario of WiMAX is using silver class of supported service application of OPNET modular and is very effective for real time application services. Figure 4 represents simulation parameters for simulation of above stated scenario:

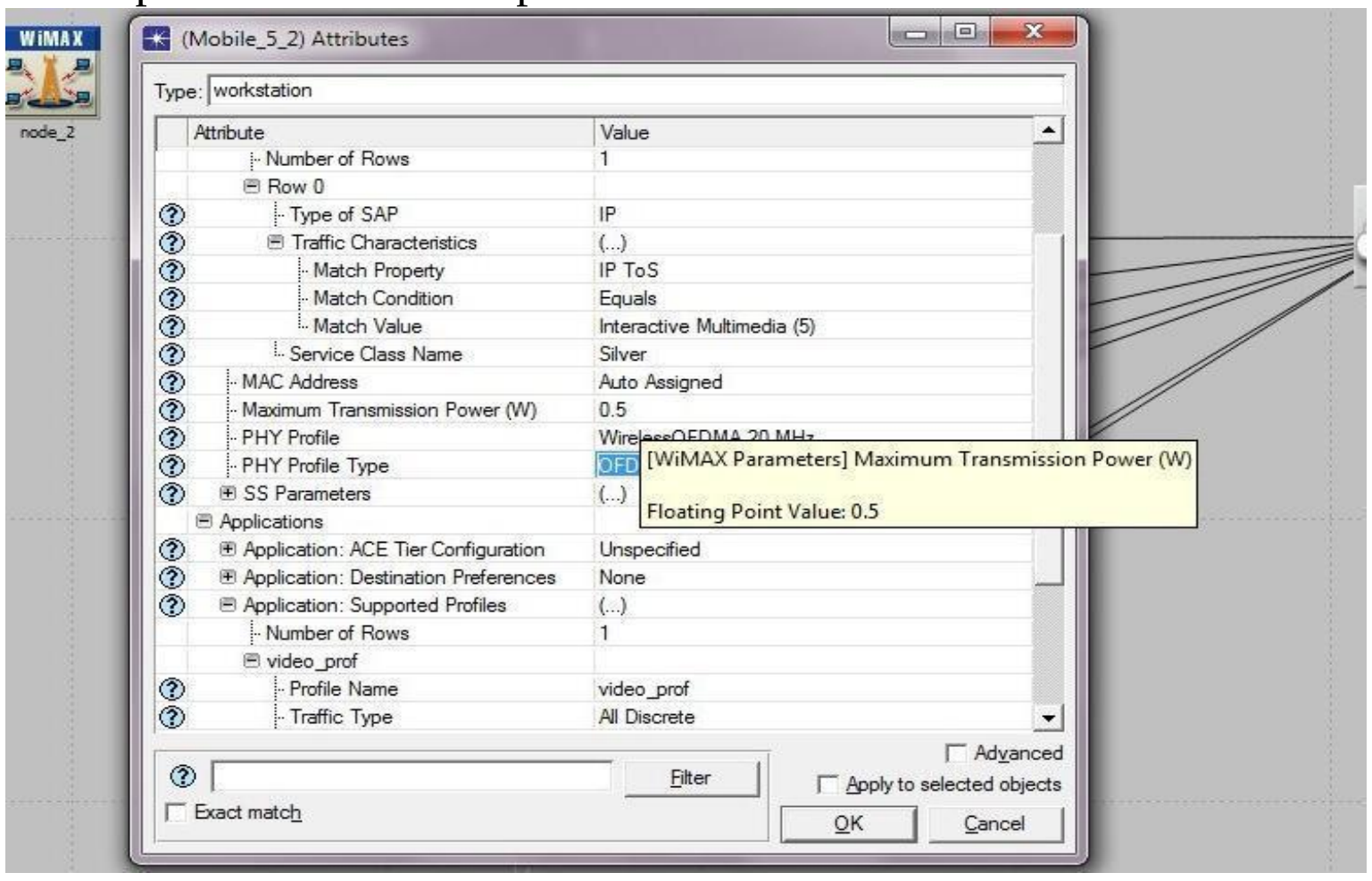

Fig 4: Simulation parameters of WiMAX using OPNET

## VI. SIMULATION RESULTS AND COMPARISION

In this paper, our goal is to propose an uplink scheduler which does efficient and fair allocation of the bandwidth. We first implement an EDF scheduler with the SSBPF approach of giving priority to each subscriber station and simulate WiMAX networks with 7 hexagonal cells. Further we will proceed to implementation of the heuristic EDF uplink scheduler (H-EDF) by using the same scenario of a WiMAX network [17]. Then in next we will compare performance of H-EDF uplink scheduler with EDF scheduler. The performance of both the scenario will be evaluated in terms of throughput and end to end delay.

- **THROUGHPUT:**

Throughput is represented in bits per second (bps) and it is the number of packets which is received in per unit of time. Figure 5 shows the throughput of EDF Scheduler using SSBPF approach for WiMAX communication network.

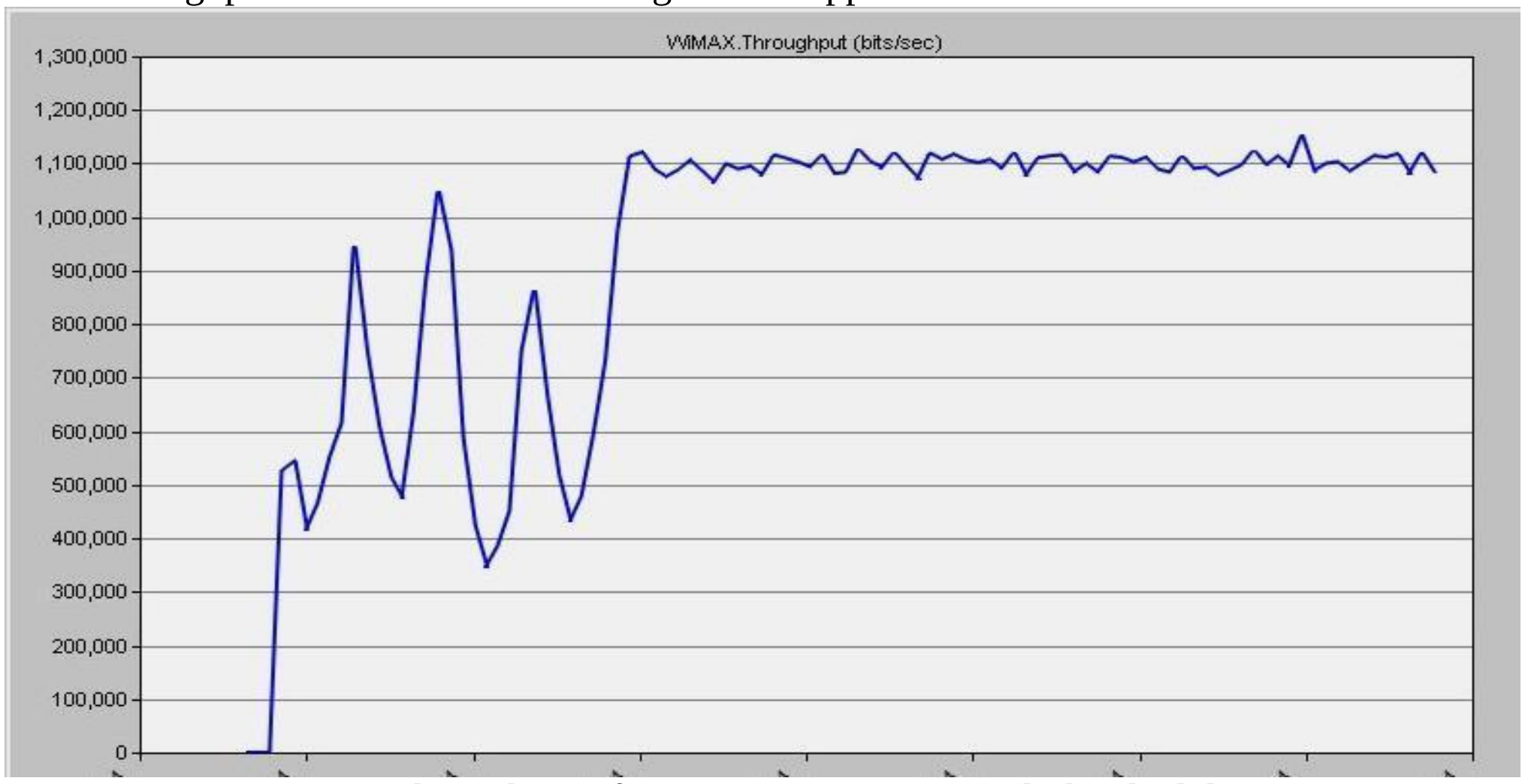

Fig 5 : Throughput of Wimax using SSBPF uplink scheduler

Followings figure 6 represents graph of proposed H-EDF uplink scheduler for WiMAX Communication system.

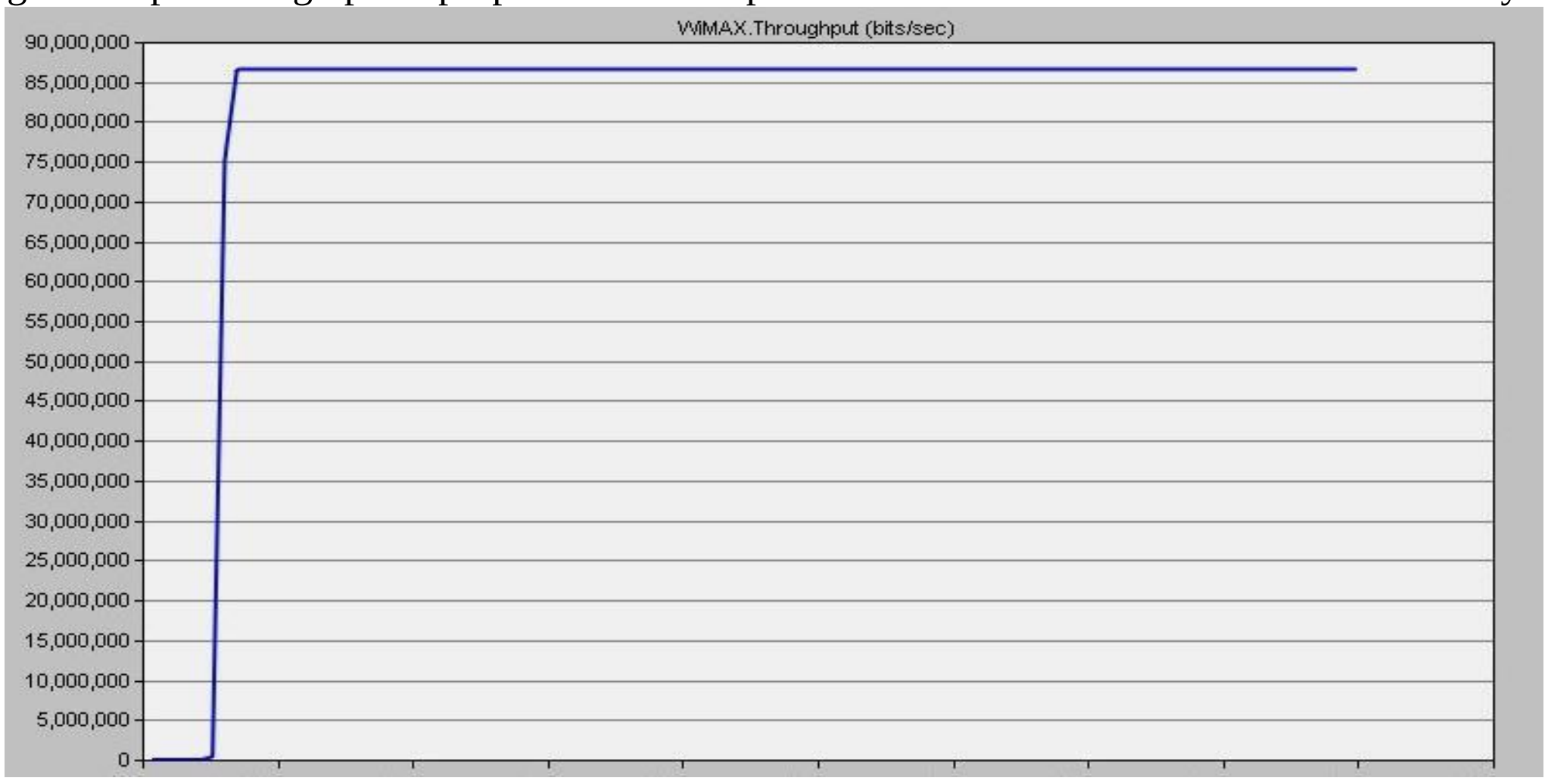

Fig 6: Throughput of WiMAX system using H-EDF uplink scheduler

- **DELAY**

End to end delay is the quantity of the time which is occupied to sending packet from source to their respective destination until receive those packets. Figure 7 represents end to end delay of WiMAX using SSBPF scheduling algorithm.

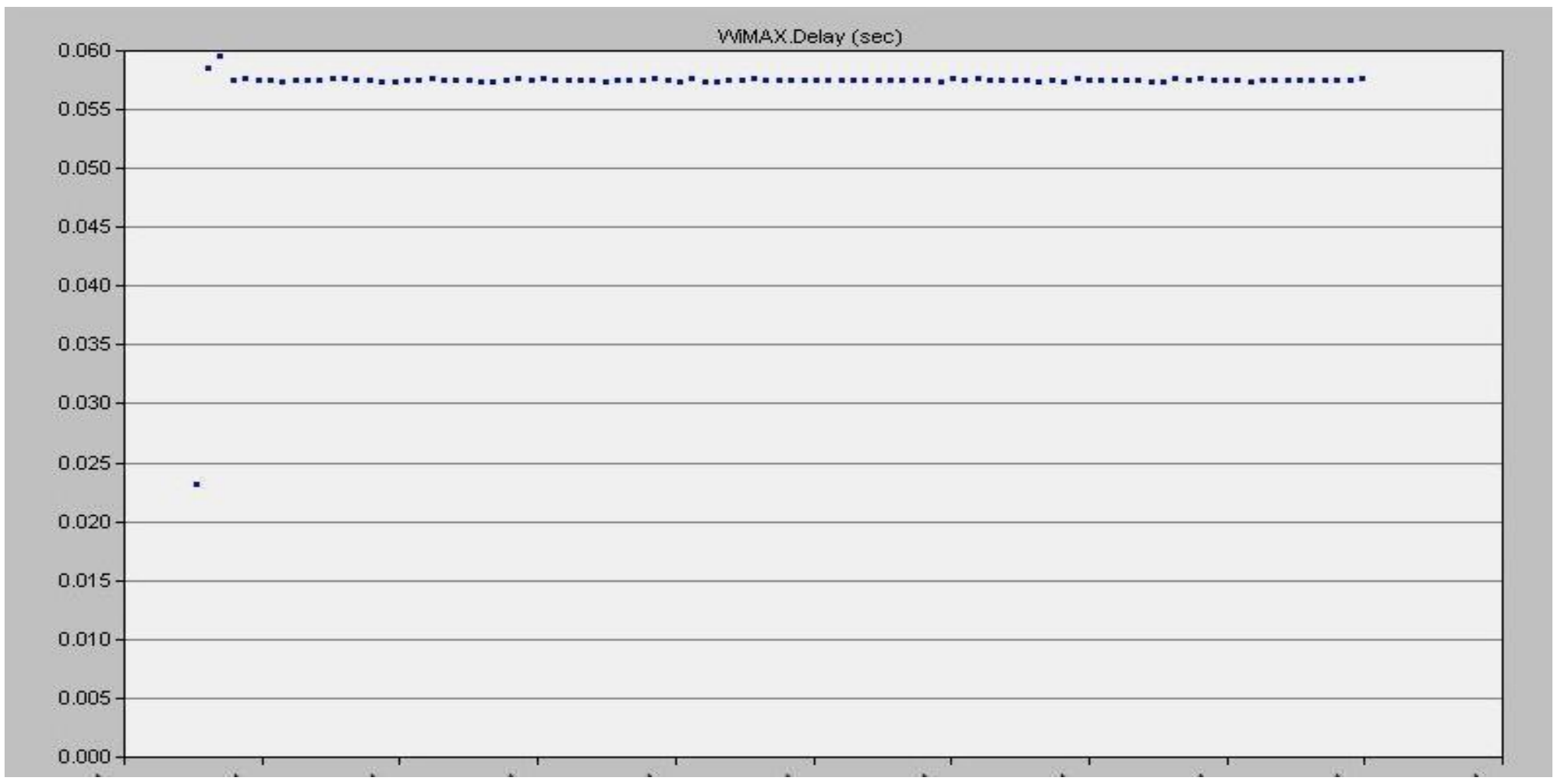

Fig 7: End to end delay of WiMAX using SSBPF scheduling algorithm

The following figure 8 represents end to end delay of WiMAX communication system which uses H-EDF uplink Scheduler:

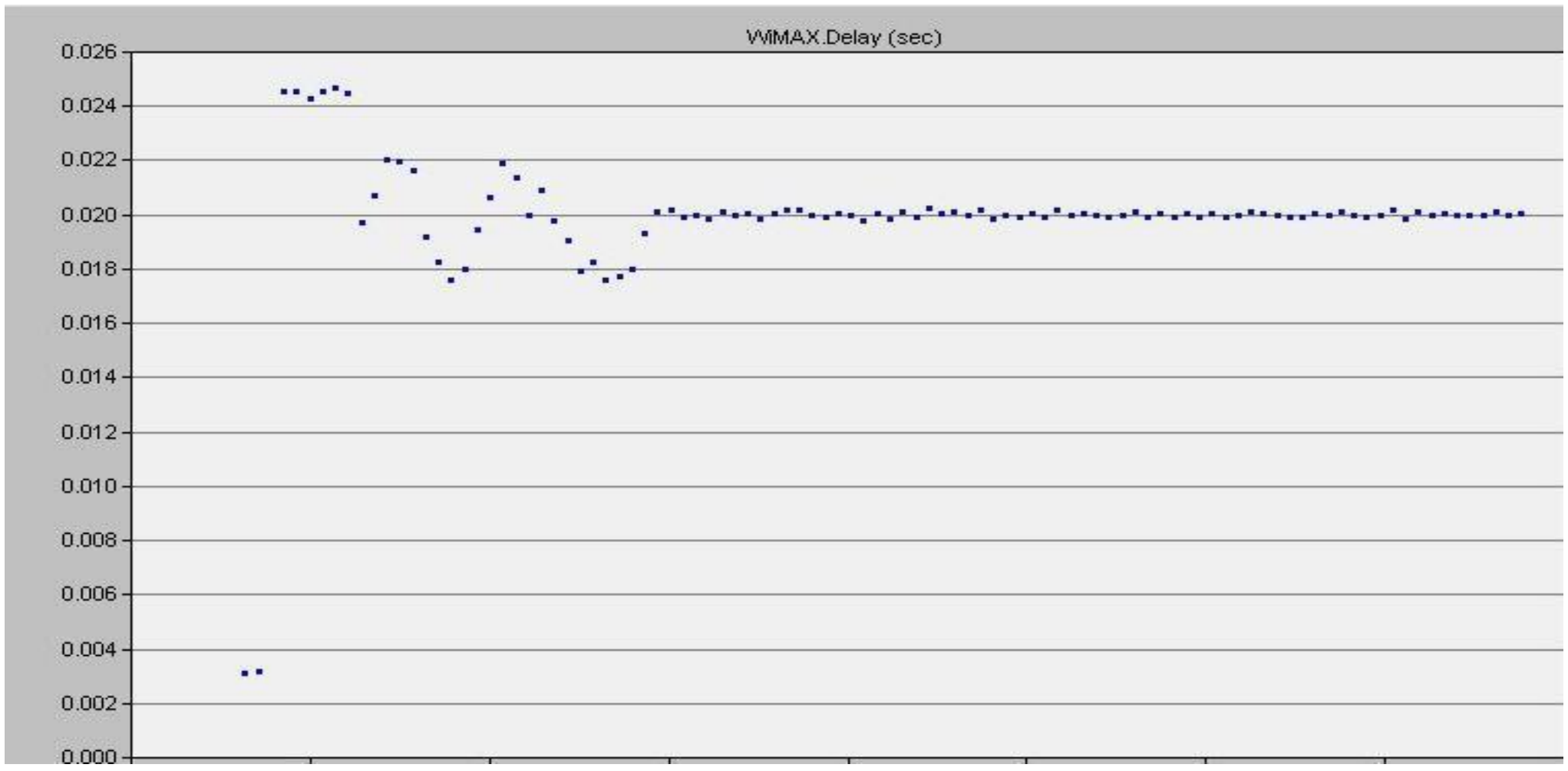

Fig 8: End to end delay of WiMAX system using H-EDF uplink scheduler

## VII. CONCLUSION

WiMAX is the most emerging technology in nowadays, it supports the IEEE 802.16 family of standards. It has very wide range and bandwidth is about 100 Mbps. It has a great feature of interoperability which supports many times of wired and wireless device. To satisfy and complete various service requests of various customers in the real time world environment, like video, internet games, video conferencing, MPEG and all other multimedia applications which are come in the category of real time applications, high bandwidth and high speed is required to carry out the desired task. To fulfill the customer's request and requirements an uplink scheduler is necessary who assigns bandwidth to the requesting subscriber station in an efficient and fair way so that speed and performance of the network will increase rapidly. The proposed H-EDF uplink scheduler allocates bandwidth to the Subscriber station in a more efficient way, hence the performance of the network will be increased. Hence the conclusion is that H-EDF uplink scheduler overcome so many limitations of the EDF algorithm like EDF does not give results in terms of throughput because of high mobility, EDF algorithm leads to problem of jitter and packet loss as well as it leads to starvation of some SS. Our Proposed H-EDF uplink scheduler will give great performance and avoids all the above issues related to EDF algorithm therefore the performance of the network will be enhanced Future work will be on providing security for WiMAX communication system from malicious node.